\documentclass[10pt]{article}
\pdfoutput=1
\usepackage{chicago,epic,graphpap,graphics,float,tabularx,paralist,longtable,fancyhdr,fancyvrb,bbm}
\usepackage{amsmath}
\usepackage{amstext}
\usepackage{amsbsy}
\usepackage{amsthm}
\usepackage{amsfonts}
\usepackage{amssymb}
\usepackage{amscd}
\usepackage{eucal}
\def\S{{\bf S}}
\def\m{\mu}
\def\D{\Delta}
\def\d{\delta}
\def\t{\tau}
\def\p{\pi}
\def\e{\eta}
\def\g{\gamma}
\def\eps{\varepsilon}
\def\as{\quad\text{{\rm a.s.}}}

\def\1{{\mathbbm 1}}

\def\P{{\mathbb P}}
\def\R{{\mathbb R}}
\def\eqdef{\triangleq}

\def\sumi{\sum_{i=1}^n}
\def\s{\sigma}

\begin{document}

\centerline{\Large\bf An example of short-term relative arbitrage}
\vspace{5pt} \centerline{\large Robert Fernholz\footnote{INTECH, One Palmer Square, Princeton, NJ 08542.  bob@bobfernholz.com. The author thanks Christa Cuchiero, Ioannis Karatzas, Constantinos Kardaras, Johannes Ruf, and Walter Schachermayer for their invaluable comments and suggestions regarding this research.}} \centerline{ \today}
\vspace{10pt}
\begin{abstract}
Long-term relative arbitrage exists in markets where the excess growth rate of the market portfolio is bounded away from zero. Here it is shown that under a time-homogeneity hypothesis this condition will also imply the existence of relative arbitrage over arbitrarily short intervals.
\end{abstract}
\vspace{10pt}

Suppose we have a market of stocks $X_1,\ldots,X_n$ represented by positive continuous semimartingales that satisfy
\[
d\log X_i(t)=\g_i(t)\,dt+\sum_{\nu=1}^{d}\xi_{i\nu}(t)\,dW_\nu(t),
\]
for $i=1,\ldots,n$, where $d\ge n\ge 2$, $(W_1,\ldots,W_d)$ is a $d$-dimensional Brownian motion, and the processes $\g_i$ and $\xi_{i\nu}$ are progressively measurable with respect to the underlying filtration with $\g_i$ locally integrable and $\xi_{i\nu}$ locally square-integrable. The process $X_i$ represents the total capitalization of the $i$th company, so the total capitalization of the market is $X(t)= X_1(t)+\cdots+X_n(t)$ and the {\em market weight processes} $\m_i$ are defined by $\m_i(t) = X_i(t)/X(t)$,
for $i=1,\ldots,n$. The $ij$th {\em covariance process} $\s_{ij}$ is defined by
\[
\s_{ij}(t)\eqdef \sum_{\nu=1}^d \xi_{i\nu}(t)\xi_{j\nu}(t),
\]
for $i,j=1,\ldots,n$.

A {\em portfolio} $\p$ is defined by its {\em weights} $\p_1,\ldots,\p_n$, which are bounded processes that are progressively measurable with respect to the Brownian filtration and add up to one. The {\em portfolio value process} $Z_\p$ for $\p$ satisfies
\[ 
d\log Z_\p(t) = \sumi \p_i(t)\,d\log X_i(t) + \g^*_\p(t)\,dt,\as,
\]
where the process $\g^*_\p$ defined by
\[
\g^*_\p(t)\eqdef\frac{1}{2}\Big(\sumi \p_i(t) \s_{ii}(t) - \sum_{i,j=1}^n \p_i(t)\p_j(t)\s_{ij}(t)\Big)
\]
is called the {\em excess growth rate process} for $\p$. It can be shown that if $\p_i(t)\ge 0$, for $i=1,\ldots,n$, then $\g^*_\p(t)\ge 0$, a.s. The market weights $\m_i$ define the {\em market portfolio} $\m$, and if the market portfolio value process $Z_\m$ is initialized so that $Z_\m(0)=X(0)$, then $Z_\m(t)=X(t)$ for all $t\ge0$, a.s. Since the market weights are all positive, $\g_\m^*(t)\ge0$, a.s. 
This introductory material can be found in \citeN{F:2002}.

Let $\S$ be the entropy function  defined by 
\[
\S(x)= -\sumi x_i\log x_i,
\]
for $x\in\D^n$, the unit simplex in $\R^n$. We see that $0\le\S(x)\le\log n$, where the minimum value occurs only at the corners of the simplex, and the maximum value occurs only at the point where $x_i=1/n$ for all $i$. For a constant $c\ge0$, the {\em generalized entropy function} $\S_c$ is defined by
\[
\S_c(x)= \S(x)+c,
\]
for $x\in\D^n$. It can be shown that $\S_c$ generates a portfolio $\p$ with weights
\[
\p_i(t)=\frac{c- \log \m_i(t)}{\S_c(\m(t))}\m_i(t),
\]
for $i=1,\ldots,n$, and the portfolio value process $Z_\p$ will satisfy
\begin{equation}\label{1}
d\log\big(Z_\p(t)/Z_\m(t)\big)= d\log \S_c(\m(t)) + \frac{\g^*_\m(t)}{\S_c(\m(t))}\,dt,\as
\end{equation}
(see \citeN{F:pgf}, \citeN{F:2002}, and \citeN{FK:2005}). 

\vspace{10pt}
\noindent{\bf Definition 1.} For $T>0$, there is {\em relative arbitrage}  versus the market on $[0,T]$ if there exists a portfolio $\p$ such that
\begin{align*}
\P\big[Z_\p(T)/Z_\m(T) \ge Z_\p(0)/Z_\m(0)\big] &=1,\\
\P\big[Z_\p(T)/Z_\m(T) > Z_\p(0)/Z_\m(0)\big] &>0.\
\end{align*}
If $\P\big[Z_\p(T)/Z_\m(T) > Z_\p(0)/Z_\m(0)\big]=1$, then this relative arbitrage is {\em strong.}

\vspace{10pt}
\noindent{\bf Proposition 1.} {\em For $T>0$, suppose that for the market $X_1,\ldots,X_n$ there exists a constant $\eps>0$ such that
\[
\g^*_\m(t)>\eps,\as,
\]
 for all $t\in[0,T]$, and for the entropy function $\S$
\begin{equation}\label{0}
 \text{\rm ess inf}\big\{\S(\m(t)): t\in[0,T/2]\big\}\le \text{\rm ess inf}\big\{\S(\m(t)): t\in[T/2,T]\big\}.
\end{equation}
Then there is relative arbitrage versus the market  on $[0,T]$.}

\vspace{10pt}
\noindent{\em Proof.} Let
\begin{equation}\label{00}
A= \text{\rm ess inf}\big\{\S(\m(t)): t\in[0,T/2]\big\}.
\end{equation}
Since $\g^*_\m(t)\ge\eps>0$ on $[0,T]$, a.s., not all the $\m_i$ can be constantly equal to $1/n$, so
\[
0\le A <\log n.
\]
Hence, we can choose  $\d>0$ such that $A+2\d<\log n$ and
\[
\P\big[\inf_{t\in[0,T/2]} \S(\m(t))<A+\d\big]>0,
\]
so if we define the stopping time
\[
\t_1= \inf\big\{t\in[0,T/2]: \S(\m(t))\le A+\d\big\}\wedge T,
\]
then 
\[
\P\big[\t_1\le T/2\big]>0.
\]
We can now define a second stopping time
\[
\t_2= \inf\big\{t\in[\t_1,T]: \S(\m(t))= A+2\d\big\}\wedge T,
\]
and we have $\t_1\le\t_2$, a.s.

Now consider the generalized entropy function
\[
\S_\d(x)\eqdef \S(x)+\d,
\]
for  the same $\d>0$ as we chose above, so $\S_\d(x)\ge\d$. It follows from \eqref{1} that
\begin{equation}\label{4} 
\log\big(Z_\p(\t_2)/Z_\m(\t_2)\big)-\log\big(Z_\p(\t_1)/Z_\m(\t_1)\big) = \log\S_\d(\m(\t_2))-\log\S_\d(\m(\t_1))+\int_{\t_1}^{\t_2}\frac{\g^*_\m(t)}{\S_\d(\m(t))}\,dt,\as,
\end{equation}
for the times $\t_1$ and $\t_2$. Suppose we are on the set where $\t_1\le T/2$, so $\t_1<\t_2$, a.s., and consider two cases:
\begin{enumerate}
\item If $\t_2<T$, then 
\[
 \log\S_\d(\m(\t_2))-\log\S_\d(\m(\t_1))\ge\log (A+3\d)-\log(A+2\d)>0,\as,
 \]
 and since the integral in \eqref{4} is positive, a.s., we have
 \begin{equation}\label{5}
\log\big(Z_\p(\t_2)/Z_\m(\t_2)\big)-\log\big(Z_\p(\t_1)/Z_\m(\t_1)\big)>0,\as
\end{equation}
\item  If $\t_2=T$, then $A+\d\le\S_\d(\m(t))<A+3\d$ for $t\in[\t_1,T]$, a.s., so
\begin{equation}\label{5.1}
 \log\S_\d(\m(\t_2))-\log\S_\d(\m(\t_1))+\int_{\t_1}^{\t_2}\frac{\g^*_\m(t)}{\S_\d(\m(t))}\,dt> \log \frac{A+\d}{A+2\d} + \frac{\eps T}{2(A+3\d)},\as
 \end{equation}
 Again there are two cases:
 \begin{enumerate}
\item If $A=0$, let 
\begin{equation}\label{6}
\d=\frac{\eps T}{6\log 2},
\end{equation}
so the left-hand side of the inequality in \eqref{5.1} will be positive, a.s., and \eqref{4} implies that 
 \begin{equation}\label{7}
\log\big(Z_\p(\t_2)/Z_\m(\t_2)\big)-\log\big(Z_\p(\t_1)/Z_\m(\t_1)\big)>0,\as
\end{equation}
\item If $A>0$, then
\begin{equation}\label{6.1}
\lim_{\d\downarrow 0} \bigg[ \log \frac{A+\d}{A+2\d} + \frac{\eps T}{2(A+3\d)}\bigg]= \frac{\eps T}{2A} >0,
\end{equation}
so for small enough $\d>0$, \eqref{5.1} will be positive, and  \eqref{7} will be valid.
\end{enumerate}
\end{enumerate}

Now consider the portfolio $\e$ defined by: 
\begin{enumerate}
\item For $t\in[0,\t_1)$, $\e(t)=\m(t)$, the market portfolio.
\item For $t\in[\t_1,\t_2)$, $\e(t)=\p(t)$, the portfolio generated by $\S_\d$ with $\d$ chosen according to  \eqref{6} or \eqref{6.1}, as the case may be.
\item For $t\in[\t_2 , T]$, $\e(t)=\m(t)$.
\end{enumerate}
If $\t_1=T$, then $\e(t)=\m(t)$ for all $t\in[0,T]$, so 
\[
\log\big(Z_\e(T)/Z_\m(T)\big)=\log\big(Z_\e(0)/Z_\m(0)\big),\as
\]
If $\t_1\ne T$, then $\t_1\le T/2$ and $\t_1<\t_2$, a.s. By the construction of $\e$, we have
\begin{align*}
\log\big(Z_\e(T)/Z_\m(T)\big)-\log\big(Z_\e(0)/Z_\m(0)\big)&=\log\big(Z_\p(\t_2)/Z_\m(\t_2)\big)-\log\big(Z_\p(\t_1)/Z_\m(\t_1)\big)\\
&>0,\as,
\end{align*}
with the inequality following from \eqref{5} or \eqref{7}, as the case may be.  Since $\P[\t_1\ne T]>0$,
\begin{align*}
\P\big[ \log\big(Z_\e(T)/Z_\m(T)\big)&\ge\log\big(Z_\e(0)/Z_\m(0)\big)\big]=1,\\
\P\big[ \log\big(Z_\e(T)/Z_\m(T)\big)&>\log\big(Z_\e(0)/Z_\m(0)\big)\big]>0,
\end{align*}
so there is relative arbitrage versus the market on $[0,T]$.\qed

\vspace{10pt}
Let us recall that the market is {\em diverse} over the interval $[0,T]$ if there exists a $\d >0$ such that
\[
\m_i(t) < 1-\d,\as,
\]
for $i=1,\ldots,n$ and all $t\in[0,T]$ (see, e.g., \citeN{F:2002}).

\vspace{10pt}
\noindent{\bf Corollary 1.}  {\em Let $T>0$ and suppose that the market is not diverse over  $[0,T/2]$ and that $\g^*_\m(t)>\eps>0$ for $t\in[0,T]$.  Then there is relative arbitrage versus the market  on $[0,T]$.} 

\vspace{10pt}
\noindent{\em Proof.} In this case $A=0$ in \eqref{00}.\qed

\vspace{10pt}
\noindent{\bf Remark 1.} Corollary~1 can be applied to  {\em volatility-stabilized} markets, for which \citeN{Banner/Fernholz:2008} have previously shown the existence of short-term strong relative arbitrage.

\vspace{10pt}
\noindent{\bf Remark 2.}  The condition \eqref{0}  can be generalized to  a function $A$ defined on $[0,T]$ by
\[
A(t)= \text{\rm ess inf}\big\{\S(\m(t))\big\}.
\]
If $A$ increases over any subinterval of $[0,T]$, then an argument similar to that of case~1 in Proposition~1 will establish relative arbitrage. Moreover,
Johannes Ruf has pointed out that the proof of Proposition~1 can be extended to establish relative arbitrage in the case where $A$ is slowly (enough) decreasing on $[0,T]$. By means of a remarkable construction,  \citeN{Karatzas/Ruf:2015} have shown that short-term relative arbitrage does not exist for arbitrary $A$.

\bibliographystyle{chicago}
\bibliography{math}
\end{document}